\documentclass{article}
\usepackage{ieeetrantools}
\usepackage{cite} 
\usepackage{spconf,amsmath,graphicx}
\usepackage{tabularx}
\usepackage[ruled,linesnumbered]{algorithm2e}
\usepackage[labelfont=bf]{caption} 
\usepackage{hyperref,url}
\usepackage{multirow}
\usepackage{booktabs}
\usepackage{amssymb}

\title{Target Sound Extraction with Variable Cross-modality Clues}
\name{
\begin{tabular}{c}
Chenda Li$^{1,2,\dag}$, Yao Qian$^2$, Zhuo Chen$^2$, Dongmei Wang$^2$, \\
Takuya Yoshioka$^2$, Shujie Liu$^2$, Yanmin Qian$^1$, Michael Zeng$^2$
\end{tabular}  \thanks{$^{\dag}$The first author conducted this work during internship at Microsoft.}}

\address{
  $^1$MoE Key Lab of Artificial Intelligence, AI Institute\\
$^1$X-LANCE Lab, Department of Computer Science and Engineering, Shanghai Jiao Tong University\\
  $^2$Microsoft, Redmond, WA, USA\\
  }

\begin{document}
\bstctlcite{IEEEexample:BSTcontrol} 
\ninept
\maketitle
\begin{abstract}

Automatic target sound extraction (TSE) is a machine learning approach to mimic the human auditory perception capability of attending to a sound source of interest from a mixture of sources. 
It often uses a model conditioned on a fixed form of target sound clues, such as a sound class label, which limits the ways in which users can interact with the model to specify the target sounds.
To leverage variable number of clues cross modalities available in the inference phase, including a video, a sound event class, and a text caption, we propose a unified transformer-based TSE model architecture, where a multi-clue attention module integrates all the clues across the modalities. Since there is no off-the-shelf benchmark to evaluate our proposed approach, we build a dataset \footnote{Python scripts to generate this dataset can be found at \url{https://github.com/LiChenda/Multi-clue-TSE-data}.} based on public corpora, Audioset and AudioCaps. Experimental results for seen and unseen target-sound evaluation sets show that our proposed TSE model can effectively deal with a varying number of clues which improves the TSE performance and robustness against partially compromised clues. 
\end{abstract}
\section{Introduction}

People can focus their auditory attention on the sound of their interest in a complex acoustic environment \cite{cherryExperimentsRecognitionSpeech1953}.
Researchers have attempted to endow machines with a similar capability by audio source separation, a process of separating all audio sources out of their mixture.
Audio source separation includes speech separation \cite{hersheyDeepClusteringDiscriminative2016,kolbaekMultitalkerSpeechSeparation2017,luoConvTasNetSurpassingIdeal2019}, music separation \cite{liuVoiceAccompanimentSeparation2020,liuChannelWiseSubbandInput2020,defossezMusicSourceSeparation2021}, and universal sound separation \cite{kavalerovUniversalSoundSeparation2019,tzinisImprovingUniversalSound2020,kongSourceSeparationWeakly2020}.

In some cases, instead of separating all sound sources, we may only be interested in a specific source in the mixed signal. 
With target sound extraction (TSE), only the sound of interest is extracted from the audio mixture given a target clue. The clues for the TSE systems can be provided in various forms.
Sound-related clues include a sound tag \cite{kongSourceSeparationWeakly2020,liCategoryAdaptedSoundEvent2022}, a reference speech signal and a target speaker embedding \cite{delcroixSingleChannelTarget2018,wangVoiceFilterTargetedVoice2019,wangSpeechSeparationUsing2019,xuSpExMultiScaleTime2020,delcroixImprovingSpeakerDiscrimination2020}. 
Visual information can also be used as the extraction clue in multi-modal target sound extraction \cite{afourasConversationDeepAudioVisual2018,gaoVisualVoiceAudioVisualSpeech2021,tzinisAudioScopeV2AudioVisualAttention2022a}. Some recent works use natural language descriptions as the clues~\cite{kilgourTextDrivenSeparationArbitrary2022,liuSeparateWhatYou2022b}. 
In \cite{ohishiConceptBeamConceptDriven2022a}, the authors use `concept' as the clue for target speech extraction, where the concept can be extracted from an image or a speech signal that is related to the target speech.

While there are various TSE systems each of which handles a specific clue form, there is still room for improvement in practical applications.
First, a single clue may be insufficient to describe a specific target sound. 
Secondly, the single-clue TSE system is not robust against device failures. For example, the single-clue vision-based TSE system may become incapacitated when there is a camera failure.
Lastly, the pieces of information provided by multiple types of clues may be complementary to each other to create a more comprehensive clue about the target sound, which could  result in extraction performance improvement.
Some recent works took advantage of multiple clues for target speech extraction \cite{liListenWatchUnderstand2020,tanAudioVisualSpeechSeparation2020,liVCSETimeDomainVisualContextual2022,rahimiReadingListenCocktail2022} and showed the performance superiority to single-clue systems.
However, most of these systems, except for \cite{rahimiReadingListenCocktail2022}, require all the clues to be available during the inference.
Also, they were built for speech signals and did not cope with general non-speech sounds.

In this paper, we propose a unified TSE system that can extract the target sound by flexibly combining multiple clues from different modalities that are available at test time, including a sound event tag, a text description, and a video clip related to the target sound. 
Designing such a system gives rise to several challenges.
First, clues of different modalities take different input forms, requiring a unified approach to processing them in the embedding space.
Secondly, some of the clues (e.g. the sound event tag) provide static information about the sound to be extracted, while 
others (e.g. the video clip) provide dynamic information which changes over time.
It is important to appropriately deal with the alignment between the clues and the input audio features.
Thirdly, the multi-clue TSE system should be able to deal with various number of input clues.  
To solve these challenges, we design a clue processing network based on multi-clue attention.
The basic idea is inspired by \cite{rahimiReadingListenCocktail2022}, which propose a unified Transformer-based model for speech extraction with text and video clues. 
In our TSE system, the observed sound mixture and all the input clues are firstly encoded into a unified embedding space with the corresponding modality encoders. Then, the encoded sound mixture can attend to all the different clues at the same time to synthesize a cross-modality clue.
This process does not require all the clues to be present, and the alignment problem is handled with an attention mechanism.
The contributions of this paper are as follows. (1) We propose a TSE model based on a multi-clue attention module to leverage a variable number of clues with different modalities. 
(2) The system robustness and the details of the attention mechanism are experimentally analyzed. (3) We build a multi-modal TSE dataset based on  public corpora, Audioset \cite{gemmekeAudioSetOntology2017} and AudioCaps \cite{kimAudioCapsGeneratingCaptions2019}.

\begin{figure*}[t]

\centering
\includegraphics[width=0.8\linewidth]{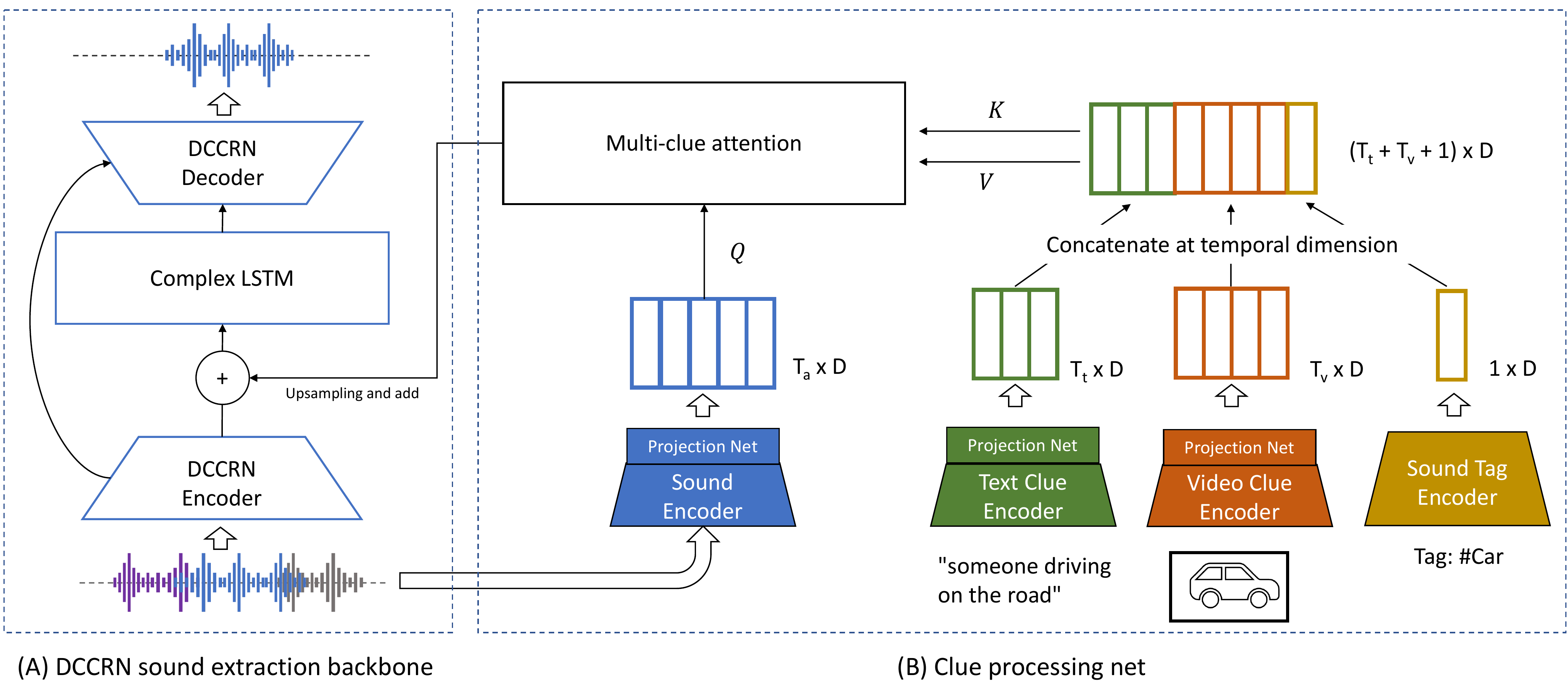}
\caption{Proposed multi-clue target sound extraction network.}
\label{fig:fig1}

\end{figure*}
\section{Target Sound Extraction}
\label{sec:baseline}

The goal of target sound extraction (TSE) is to extract the sound of interest from an audio mixture given a set of one or more target clues.
Let $\mathbf{y}$ denote the input audio mixture consisting of $J$ sound sources $\mathbf{s}_{1}, \cdots, \mathbf{s}_{J}$, and suppose the $j$-th source to be the target sound. 
Then, the TSE mapping function can be formulated as
\begin{align}
		f_{TSE}(\mathbf{y}, \mathbf{c}_{j}) = \hat{\mathbf{s}}_{j} \rightarrow \mathbf{s}_{j}, 
\end{align}
where $\hat{\mathbf{s}}_{j}$ is the estimated target sound and $\mathbf{c}_j$ is a target sound representation from the provided clue set. 
One of the simplest forms of the clue set contains only one sound event tag represented as a one-hot vector \cite{kongSourceSeparationWeakly2020,liCategoryAdaptedSoundEvent2022}, and it is used as our baseline system. In our proposed multi-clue system, $\mathbf{c}_{j}$ is produced by a multi-clue processing net as described in Section \ref{sec:multi-clue-net}.

The backbone of our TSE system is based on a deep complex convolution recurrent network (DCCRN) \cite{huDCCRNDeepComplex2020}. It was originally proposed for speech enhancement using complex time-frequency spectra and also applied to target speech extraction~\cite{9746962}. As Fig. \ref{fig:fig1} (A) shows, the DCCRN consists of an encoder, an enhancement LSTM, and a decoder. The encoder and decoder consist of complex convolution layers and are connected with U-Net-like  skip-connections~\cite{ronnebergerUNetConvolutionalNetworks2015,stollerWaveUNetMultiScaleNeural2018}. The enhancement LSTM between the encoder and the decoder processes the sum of the complex deep encoded features and the clue embedding features as follows: 
\begin{align}
 \label{eq:dccrn_1}
 \mathbf{F}_{rr} &= \text{LSTM}_{r}(\mathbf{Y}_r + \mathbf{c}_{j}), \quad \mathbf{F}_{ir} = \text{LSTM}_{r}(\mathbf{Y}_i + \mathbf{c}_{j}),  \\
\label{eq:dccrn_2}
 \mathbf{F}_{ri} &= \text{LSTM}_{i}(\mathbf{Y}_r + \mathbf{c}_{j}), 
 \quad \mathbf{F}_{ii} = \text{LSTM}_{i}(\mathbf{Y}_i + \mathbf{c}_{j}), \\
  \mathbf{F}_{out} & = ( \mathbf{F}_{rr} -  \mathbf{F}_{ii}) + ( \mathbf{F}_{ri} +  \mathbf{F}_{ir})i, 
\end{align}
where $\mathbf{Y}_{r}, \mathbf{Y}_{i} \in \mathbb{R} ^{T \times D} $ are the real and imaginary parts of the complex deep features generated by the DCCRN encoder with $T$ and $D$ being the feature sequence length and dimension, respectively. $\mathbf{c}_j \in \mathbb{R} ^{T \times D} $ is the encoded clue of the target sound, which is mapped from the one-hot tag vector with a linear layer and tiled to length $T$. $\text{LSTM}_{r},\text{LSTM}_{i}$ are LSTMs for the real and imaginary features.  $\mathbf{F}_{out}$ is the complex features that are fed to the DCCRN decoder.
The DCCRN decoder maps $\mathbf{F}_{out}$ into the estimate target sound $\hat{\mathbf{s}}_{j}$. 

The model is trained to minimize the following loss function: 
\begin{align}
\label{eq:loss}
	\mathcal{L} &= \mathcal{L}_{snr} + \lambda \mathcal{L}_{L1}, \\
	\mathcal{L}_{snr} &= -10 \log_{10}{\left ( \frac{\|\mathbf{s}_{j}\|_2^{2}}{\|\mathbf{s}_j - \hat{\mathbf{s}}_j \|_2^{2} } \right )},  \\
	\mathcal{L}_{L1} &= \|\mathbf{S}_j - \hat{\mathbf{S}}_j \|_1,
\end{align}
where $\mathbf{S}_j$ and $\hat{\mathbf{S}}_j$ are the complex spectra of the target and estimated sources, respectively, and $\lambda$ was set at 5 in our experiments.

\section{Multi-clue Processing Net}
\label{sec:multi-clue-net}

Fig. \ref{fig:fig1} shows a diagram of the proposed multi-clue TSE model.
It consists of two modules: a DCCRN TSE backbone (Fig. \ref{fig:fig1} (A)) and a multi-clue processing net (Fig. \ref{fig:fig1} (B)), where the backbone model is the same as the one described in the previous section. 
The clue processing net takes a variable number of clues as input and generates a time-aligned fused clue, which is fed to DCCRN's LSTM block as $\mathbf{c}_j$ in Eqs. \eqref{eq:dccrn_1} and \eqref{eq:dccrn_2}.
To deal with clues from different modalities, the clue processing net uses different modal encoders, including a sound encoder, a text clue encoder, a video clue encoder, and a sound tag encoder, which transform the corresponding clue inputs to a unified $D$-dimensional space. 
On top of the modal encoders, we use an attention module \cite{vaswaniAttentionAllYou2017} for clue fusion.

\noindent \textbf{Sound encoder}: The sound encoder extracts $D$-dimensional sound embeddings from the input mixture signal.
We adopt a pre-trained sound event detection (SED) model \cite{kongPANNsLargeScalePretrained2020} as the sound encoder to capture the sound event discriminated representation.
The original SED model is pre-trained for an audio classification task and outputs an SED probability mass over the audio classes for an entire input audio clip.
To keep dynamic time-dependent information, we use the frame-wise hidden embeddings obtained before the temporal aggregation layer of the SED model.
Then, we use a trainable modality projection net, to obtain a $D$-dimensional sound embedding sequence $\mathbf{Q} \in \mathbb{R}^{T_{a} \times D}$, where $T_{a}$ is the sound embedding sequence length.
The modality projection net comprises a layer-norm module and a fully connected layer with ReLU activation. Modality projection nets of this structure are also used for the text and video clues to map the text and video embeddings to the $D$-dimensional unified space. 

\noindent \textbf{Text clue encoder}: The text clue is a natural language description about the target sound. The function of the text encoder is to transform the target sound sentence into deep embeddings that can be used as the TSE clue.
We adopt a pre-trained DistilBERT \cite{sanhDistilBERTDistilledVersion2019} as our text encoder, where DistilBERT is a self-supervised learning (SSL) model trained on large-scale text data.
Token-level hidden embeddings are extracted with the pre-trained DistilBERT and then mapped by a trainable modality projection net to obtain $D$-dimensional text clue embedding $\mathbf{O} \in \mathbb{R}^{T_{t} \times D}$ in the unified space, where $T_t$ is the number of word tokens in the sentence.

\noindent \textbf{Video clue encoder}: The video clue is based on a video clip related to the target sound. As with the text clue encoder, we use an SSL model based on Swin Transformer \cite{liuSwinTransformerHierarchical2021} that is pre-trained on large-scale image data as our video clue encoder. Each image frame in the video clip is processed by the pre-trained Swin Transformer, and its output is mapped into the unified embedding space with a learnable modality projection net. The video clue embedding sequence is denoted as $\mathbf{V} \in \mathbb{R}^{T_{v} \times D}$, where $T_{v}$ is the number of the image frames in the video clip.

\noindent \textbf{Sound tag encoder}: The sound tag encoder is a simple linear layer, it takes a one-hot sound event tag as input and outputs an embedding vector $\mathbf{E} \in \mathbb{R}^{1 \times D}$.

\noindent \textbf{Multi-Clue Attention}: The sound tag encoder fuses the embeddings from the text clue, video clue, and sound tag encoders to generate a clue sequence alinged with the embeddings from the sound encoder. 
We achieve this by using source-target attention \cite{vaswaniAttentionAllYou2017}, where the queries are extracted from the sound encoder output $\mathbf{Q}$ while the key-value pairs are obtained from the concatenated embeddings. 
Specifically, we concatenate the embedded clues \footnote{When one of the clues, for example, the video clue is missing, the video encoder obtains no input and outputs nothing. Then, Eq. \ref{eq:cat} will be $	\mathbf{U} = \text{Concatenate}(\mathbf{O}; \mathbf{E}) \in \mathbb{R}^{(T_{t} + 1) \times D}$.} of different modalities along the sequence dimension to obtain the concatenated multi-modal clue $\mathbf{U}$ in the unified embedding space, i.e., 
\begin{align}
\label{eq:cat}
	\mathbf{U} = \text{Concatenate}(\mathbf{O};\mathbf{V}; \mathbf{E}) \in \mathbb{R}^{(T_{t} + T_{v} + 1) \times D}.
\end{align} 

By using the sound embedding sequence, $\mathbf{Q}$, and $\mathbf{U}$ as the query and key-value pairs in the attention module respectively, we can get fused clue $\mathbf{C}_{u}$ as follows:
\begin{align}
	\mathbf{C}_{u} &= \text{MultiHeadAttention}(\mathbf{Q}, \mathbf{U},\mathbf{U}) \in \mathbb{R}^{T_{a} \times D}, 
\end{align}
where $\text{MultiHeadAttention}(Q, K, V )$ represents a multi-head attention \cite{vaswaniAttentionAllYou2017} with query $Q$, key $K$, and value $V$.
We can see that the fused clue $\mathbf{C}_{u}$ have the same length, $T_{a}$, as the sound embedding $\mathbf{Q}$. 
To match the sequence length ($T_a$) to that of the DCCRN encoder $T$, $\mathbf{C}_{u}$ is up-sampled before being added to 
$\mathbf{Y}$ in Eqs. \eqref{eq:dccrn_1} and \eqref{eq:dccrn_2}.

\begin{table}[]
\caption{SNRi (dB) for  seen and unseen test sets. $\checkmark$: clue is available in inference stage.}
\label{tab:tab1}
\vspace{-.8em}
\centering
\begin{tabular}{l|ccc|cc}
\hline
\hline
\multicolumn{1}{c|}{\multirow{2}{*}{Model}} & \multicolumn{3}{c|}{Clues used} & \multicolumn{1}{c}{\multirow{2}{*}{Seen}} & \multicolumn{1}{c}{\multirow{2}{*}{Unseen}} \\ \cline{2-4}
\multicolumn{1}{c|}{}                       &       tag      &     text      &      video     & \multicolumn{1}{c}{}                      \\ \hline
DCCRN-tag-clue                           &       $\checkmark$     &           &           &     6.4           & 6.0                           \\ 
\hline
DCCRN-text-clue                                &            &      $\checkmark$      &           &     6.3   & 5.9                                   \\ \hline
DCCRN-video-clue                              &            &           &        $\checkmark$    &     5.9        & 5.6                              \\ \hline
\multirow{7}{*}{DCCRN-multi-clue}         &     $\checkmark$        &     $\checkmark$       &     $\checkmark$       &  \textbf{6.9} & \textbf{6.5}  \\ \cline{2-6} 
                                            &      $\checkmark$       &   $\checkmark$   &           &  6.8  & 6.4  \\ \cline{2-6}
                                            &          &     $\checkmark$        &       $\checkmark$      &  6.5  & 6.4 \\ \cline{2-6}
                                            &      $\checkmark$       &           &      $\checkmark$        &  6.6  & 6.4  \\ \cline{2-6}

                                            &      $\checkmark$       &           &           &  6.4  & 6.2 \\ \cline{2-6}
                                            &            &       $\checkmark$     &           &   6.3  & 6.0 \\\cline{2-6}
                                            &            &           &     $\checkmark$       &   5.8 & 5.9  \\
                                            \hline
\hline                                      
\end{tabular}
\end{table}
\begin{table}[]
\caption{Impacts of compromised clues on SNRi (dB). $*$: compromised clue.}
\vspace{-.8em}
\label{tab:tab2}
\centering

\begin{tabular}{ccc|cc|c}
\hline
\hline
 \multicolumn{5}{c|}{Correct/compromised clues} & \multicolumn{1}{c}{\multirow{2}{*}{SNRi}}  \\ \cline{1-5}
& text~~~  & ~~text$*$~~ &  ~video~ & ~video$*$~ & \multicolumn{1}{c}{}                      \\ \hline
  &   & $\checkmark$  &    &      &  5.8  \\  \hline
  &   & $\checkmark$  &  $\checkmark$  &      &  5.8  \\  \hline
    &   &   &    &  $\checkmark$   &  5.4 \\  \hline
  & $\checkmark$  &   &    &  $\checkmark$  & 6.3 \\  \hline
  &   &  $\checkmark$ &    &  $\checkmark$  & 6.0 \\  \hline

\hline
\hline                                      
\end{tabular}
\end{table}

\begin{figure*}[t]

\centering
\includegraphics[width=0.75\linewidth]{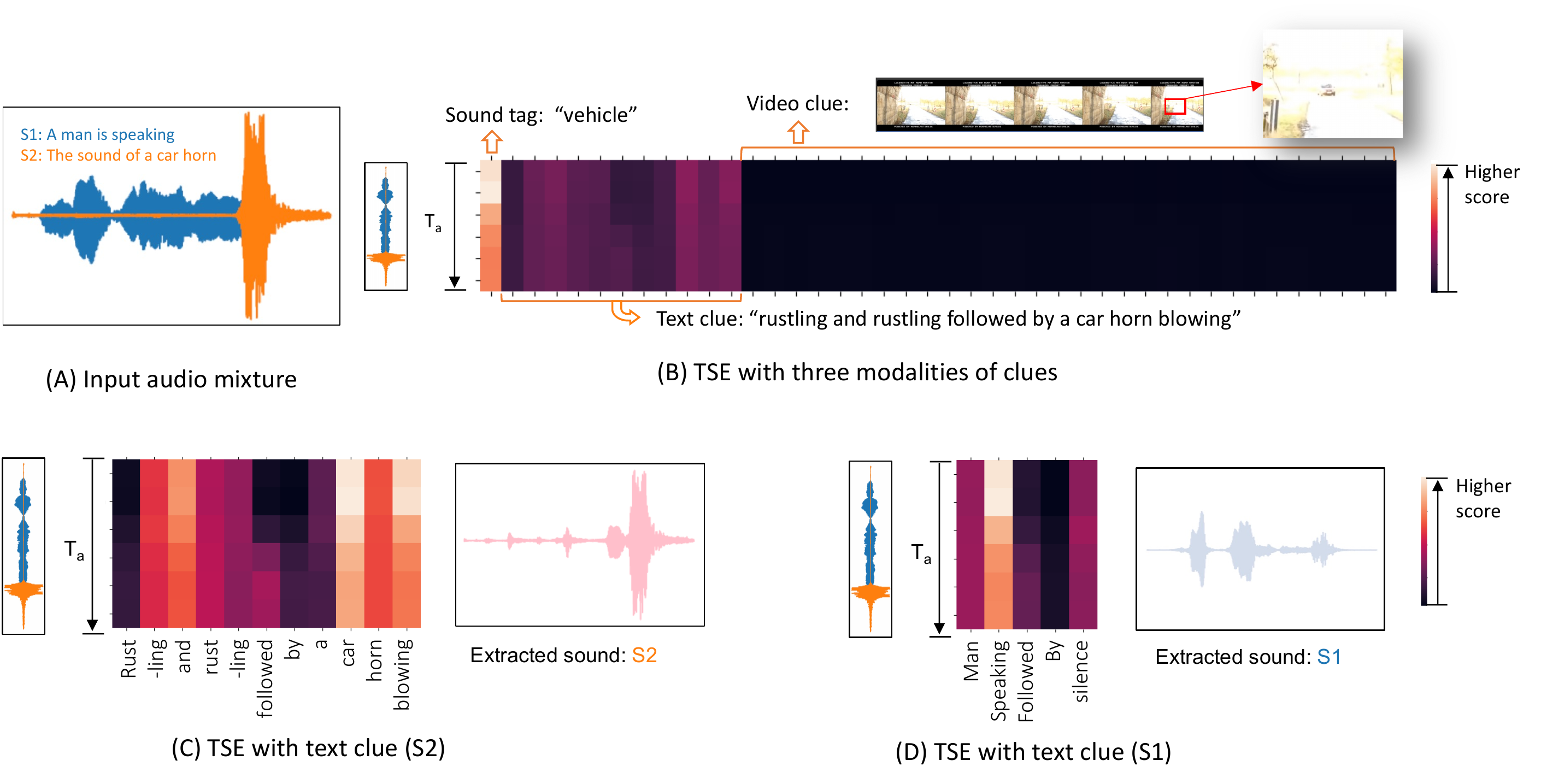}
\caption{Analysis of attention scores with different clues.}
\label{fig:fig2}
\end{figure*}
\vspace{-.9em}

\section{Experiments}
\label{sec:exp}

\subsection{Dataset}

\noindent \textbf{Audio simulation}.
We first simulate a dataset for the TSE task based on AudioSet \cite{gemmekeAudioSetOntology2017}.
AudioSet is a large-scale audio dataset drawn from YouTube videos, and it has 527 sound classes labeled by humans.
Most of the data in AudioSet are 10-second video clips with sound track.
AudioSet is a weakly labeled dataset. 
There is usually more than one sound event in a video clip with labeled tags, and the occurrence time of the sound events is not provided.
To ensure each clip has only one audio source and correct label,  the pre-processing method from  \cite{kongSourceSeparationWeakly2020} was employed in our simulation.
For each audio clip, a pre-trained SED model \cite{kongPANNsLargeScalePretrained2020} is applied to locate its sound event anchor by comparing the SED probability in 10s audio clip. The audio of 2s around the event anchor is then selected for each clip for later simulation.
For the training set, we clipped $64$k (about $35$h) sound sources of $463$ classes for simulation and simulated $124$k (about $70$h) audio mixture of two sound sources. 
The signal-to-noise ratio (SNR) of the target sound is randomly sampled between $-2$ to $2$ dB during the simulation.
For validation and testing, we simulated $0.5$h and $1$h data, respectively.
And all the target sound classes in the validation and test set are seen in the training.
Besides, we also simulated $0.7$h data of unseen target sound classes for testing, in which most of the sound classes are musical instruments.

\noindent \textbf{Text clue}. The AudioCaps \cite{kimAudioCapsGeneratingCaptions2019} provides human-written natural language description for a subset of the AudioSet. However, after the single-source clipping for the target sound, the description of the original 10s audio is no longer precise for the clipped-out 2s audio.
So, we adopt an audio caption generation model \cite{wuAudioCaptionListen2019} to create pseudo natural descriptions as the text clues.
The 2s target sound is sent to the audio caption generation model, and the model outputs a sentence that describes the target sound.

\noindent \textbf{Visual clue}. The data in the AudioSet has parallel video and audio. For the visual clue, we simply use the frames (with FPS 15) from the $2s$ video that aligned with the target sound. 

\noindent \textbf{One-hot tag clue}. As mentioned above, the AudioSet only provided weak sound class labels without the occurrence time, and we clipped out 2s audio from the original data with a SED model.
During the clipping, we transform the SED probability of the 2s audio into one-hot vectors as the clue of the target sound.

\subsection{Model configuration and training}
The FFT size, the window length, and the hop length for STFT were set to $512$, $400$, and $100$, respectively, for 16 kHz input. 
The channel numbers of the DCCRN encoder were $\{ 32, 64, 128, 256,256,256\}$.
The complex LSTM consisted of two (real and imaginary) two-layer bidirectional LSTMs with $512$ hidden units.
We used ESPNet-SE toolkit \cite{liESPnetSEEndToEndSpeech2021} for implementation. 
Open text\footnote{{https://huggingface.co/distilroberta-base}} and vision\footnote{{https://huggingface.co/microsoft/swin-large-patch4-window7-224}} encoders were used for the text  and video clues. 

The proposed multi-clue TSE system was trained  with two stages. 
In the first stage, we first trained the DCCRN backbone by using only the sound tag clue. 
In the second stage, starting from this pre-trained model, we trained all the model parameters except for the pre-trained text and video encoders to minimize the extraction loss of Eq. \eqref{eq:loss}. 
We adopted the two-stage approach because our preliminary experiment found that training the entire model from scratch in an end-to-end fashion based on Eq. \eqref{eq:loss} was hard to achieve optimal checkpoint.  
This could be because the clue processing net cannot produce stable clue embeddings in the early training steps. The two-stage training approach addresses this by starting the training with the simple sound tag-based clue. 
To enable fair comparison, the same initialization scheme was used for training the single-clue baselines.
The training was performed by using an Adam optimizer with an initial learning rate of $ 0.5 \times 10^{-4}$ and recuding it by a factor of $0.97$ for every epoch until convergence.

\subsection{Results}

\noindent\textbf{Comparison with baselines}.
Table \ref{tab:tab1} compares the proposed multi-clue TSE system with three single-clue TSE baselines for different clue-usage conditions in terms of SNR improvement (SNRi).
The proposed TSE system achieved the best SNRi score for both \textit{seen} and \textit{unseen} test sets when it used all the three clues. 
Even with two clues, our multi-clue TSE outperformed all the single-clue baselines.
When only one clue was provided,  the proposed multi-clue TSE performed comparably with the single-cue baselines for the seen test set and marginally performed better for the unseen test set. These results show that the superiority of the proposed system in terms of both effectiveness and flexibility. 

\noindent\textbf{Robustness to compromised clues}
In real applications, some of the the input clues may sometimes become inaccurate. 
To test the robustness of the proposed model, we carried out experiments with text and video clues where one or two of them were artificially compromised. 
This was done as follows. 
For the text clue, one third of the words in the clue sentence were replaced with random words. 
For the video clue, a Gaussian noise of $-2.5$dB is added to the original video clips.
Since there is no ambiguity in tag clues, we do not include the tag clues here.
Table \ref{tab:tab2} shows the experimental results. While the performance degradation was observed when one or two clues were compromised, good SNR improvements were still observed for all conditions. 
By comparing the 1st row with the 2nd row, and the 3rd row with the 4th row, we found when the text clue was compromised, adding correct video clue helps, and vice versa.
When both clues were compromised, it still showed better performance than only using one compromised clue.

\subsection{Analysis of multi-clue attention}
To analyze the attention mechanism by which the multi-clue net processes clues, we plot the attention weight matrix in figure. \ref{fig:fig2}. The input audio is mixed by two sound events, a man's speech and the sound of a car horn.
The available clues are the sound tag ``vehicle'', a natural description of the car horn, and a video in which a car is driving.
Figure \ref{fig:fig2}.b plots the attention scores in the multi-clue attention module. We can see that the highest scores are mainly in the tag and text clues. In this case, the video clues contributed little. That may be because the car in the video is too far from the camera.
When we only use the text as the extraction clue, we can see both the sound of the car horn and the sound of speech can be extracted with different descriptions as figure \ref{fig:fig2}.c and figure \ref{fig:fig2}.d) show. And the words in the sentence related to the target sound get higher scores in the attention matrix.

\vspace{-0.8em}
\section{Conclusion}
\label{sec:conclusion}

In this work, we propose a multi-clue model for target sound extraction. The proposed model can freely combine different modal clues to extract target sounds. 
The experiments show that the proposed multi-clue TSE can leverage each single clue effectively to extract target sounds, achieving comparable performance with single-clue based systems. 
When combining clues from multiple modalities, the proposed model shows further improvements in both performance and robustness.
These advantages make the proposed TSE system strong, robust, and flexible in real applications. 
With these observations, in future work, we would like to investigate further integration with more clues and their interaction with each other during TSE process.

\bibliographystyle{IEEEtran}
\bibliography{refs}

\end{document}